\documentclass[twocolumn, prb, a4paper, 10pt, superscriptaddress, nopacs, floatfix]{revtex4-1}
\usepackage{siunitx}
\usepackage{graphicx}
\usepackage{xcolor}
\usepackage{physics}
\usepackage{nicefrac}
\usepackage{float}
\usepackage[utf8]{inputenc}
\usepackage[colorlinks, citecolor=blue, linkcolor=cyan]{hyperref}
\usepackage{ulem}

\graphicspath{{Figures/}}

\begin{document}

\title{Quantitative study of the response of a single NV defect in diamond to magnetic noise}

\author{Maxime Rollo}
\author{Aurore Finco}
\author{Rana Tanos}
\author{Florentin Fabre}
\affiliation{Laboratoire Charles Coulomb, Université de Montpellier and CNRS, 34095 Montpellier, France}
%\author{Farah Amar}
%\author{Paul Crozat}
\author{Thibaut Devolder}
\affiliation{Centre de Nanosciences et de Nanotechnologies, CNRS, Université Paris-Saclay, 91120 Palaiseau, France}
\author{Isabelle Robert-Philip}
\author{Vincent Jacques}
\affiliation{Laboratoire Charles Coulomb, Université de Montpellier and CNRS, 34095 Montpellier, France}

\begin{abstract}
The nitrogen-vacancy (NV) defect in diamond is an efficient quantum sensor of randomly fluctuating signals via relaxometry measurements. In particular, the longitudinal spin relaxation of the NV defect accelerates in the presence of magnetic noise with a spectral component at its electron spin resonance frequency. We look into this effect quantitatively by applying a calibrated and tunable magnetic noise on a single NV defect. We show that an increase of the longitudinal spin relaxation rate translates into a reduction of the photoluminescence (PL) signal emitted under continuous optical illumination, which can be explained using a simplified three-level model of the NV defect. This PL quenching mechanism offers a simple, all-optical method to detect magnetic noise sources at the nanoscale.
\end{abstract} 
\date{\today}

\maketitle

Quantum sensors take advantage of the extreme sensitivity of quantum systems to external perturbations to accurately measure a broad range of physical quantities such as acceleration, rotation, magnetic and electric fields, or temperature~\cite{degenQuantumSensing2017}. Among a wide variety of quantum systems employed for sensing purposes, the nitrogen-vacancy (NV) defect in diamond has garnered considerable attention in the last decade for the development of highly sensitive magnetometers~\cite{Gupi_Nature2008,Maze2008,rondinMagnetometryNitrogenvacancyDefects2014,doi:10.1146/annurev-physchem-040513-103659}. Besides the detection of static magnetic fields, the NV defect can also be employed for magnetic noise sensing by recording variations of its longitudinal spin relaxation time~$T_1$~\cite{hall_sensing_2009}. This technique, called relaxometry, is a powerful tool for detecting the magnetic noise produced by thermal fluctuations of charges or spins in solids, which are closely linked to their intrinsic physical properties like conductivity, magnetic resonance or spin wave dispersion~\cite{casolaProbingCondensedMatter2018}. In recent years, NV-based relaxometry was successfully applied to the characterization of Johnson noise in conductors~\cite{kolkowitzProbingJohnsonNoise2015, ariyaratneNanoscaleElectricalConductivity2018}, the observation of electronic instabilities in graphene~\cite{andersenElectronphononInstabilityGraphene2019}, the detection of paramagnetic molecules~\cite{Ermakova2013,tetienneSpinRelaxometrySingle2013,Steinert2013,PhysRevApplied.2.054014,Lukin_NanoLett2014,schmid-lorchRelaxometryDephasingImaging2015,TetienneNanolett2016}, the measurement of pH and redox potential in microfluidic channels~\cite{Rendler2017,pH2019,Cigler2020}, the study of spin waves in magnetic materials~\cite{vandersarNanometrescaleProbingSpin2015, duControlLocalMeasurement2017,lee-wongNanoscaleDetectionMagnon2020} and nanoscale imaging of non-collinear spin textures in synthetic antiferromagnets~\cite{fincoImagingNoncollinearAntiferromagnetic2020}.

\begin{figure}[h!]
  \centering
  \includegraphics[scale=0.9]{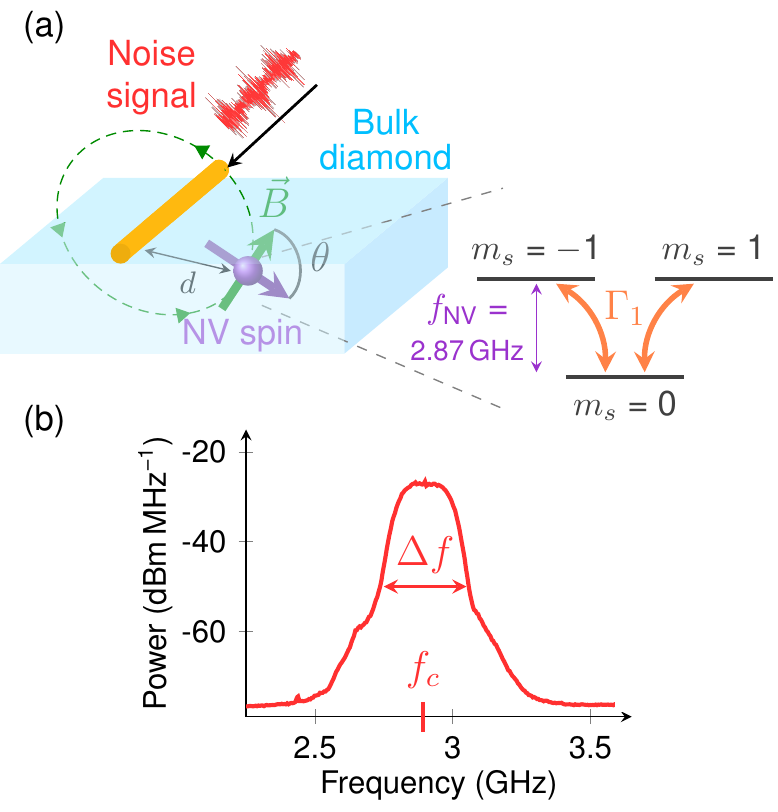}
  \caption{(a) Principle of the experiment. A calibrated noise signal is sent through a copper microwire spanned on a bulk diamond sample. This noise signal produces a fluctuating Oersted field with a tunable spectral density at the electron spin resonance frequency $f_{\rm NV}$ of a single NV defect located at a distance $d$ from the microwire. The ground state spin sublevels of the NV defect, $m_s = 0, \pm 1$, are shown on the right. The NV quantization axis forms an angle $\theta$ with the direction of the magnetic field noise. (b) Typical noise spectrum used in the experiments. }
  \label{fig:setup}
\end{figure}

Given the large range of applications of single spin relaxometry, the purpose of the present work is to provide a quantitative study of the response of a single NV defect in diamond to magnetic noise. To this end, we apply a calibrated and tunable magnetic noise on a single NV defect and measure its effect both on the longitudinal spin relaxation time $T_1$ and on the photoluminescence (PL) signal emitted under continuous optical illumination. We show that a reduction of the relaxation time $T_1$ results in an overall decrease of the PL signal, providing a practically and conceptually simple mechanism to detect magnetic noise.  The evolution of the magnetic-noise-induced PL quenching is then studied as a function of the optical pumping power. All the experimental results are well described by a simplified closed three-level model of the NV defect, from which we infer a shot noise limited sensitivity to magnetic noise $\eta_\text{cw}\sim \SI{1}{\micro\tesla\squared{\mega\Hz}^{-1}\per\sqrt{\rm Hz}}$. %We show that this value is similar to the one obtained for alternative relaxometry protocols based on pulsed laser excitation.

The principle of the experiment is sketched in Fig.~\ref{fig:setup}(a). Individual NV defects hosted in a ultrapure, bulk diamond crystal are optically isolated at room temperature using a scanning confocal microscope operating under green laser illumination. A tunable noise signal is obtained by mixing a low frequency white noise source with a microwave carrier frequency and filtering out the DC component. As shown in Fig.~\ref{fig:setup}(b), the resulting noise spectrum exhibits a roughly constant power spectral density over a frequency window of adjustable width $\Delta f$ and center position $f_\text{c}$. This noise signal is sent through a copper microwire directly spanned on the diamond surface, thus converting noise current into a fluctuating Oersted field. Assuming that the noise power is constant over the frequency window $\Delta f$, the field spectral density $S_B$ at a distance $d$ from the microwire is given by
\begin{equation}
  \label{eq:S_B}
  S_B =
  \begin{cases}
    \displaystyle \frac{\mu_0^2 P_\text{n}}{4\pi^2 d^2 R \Delta f} \ & \text{for} f \in [f_\text{c} -\nicefrac{\Delta f}{2}, f_\text{c} +\nicefrac{\Delta f}{2}]\\
    & \\
     \ \ \ \ \ \ 0 \ & \text{for} f \notin [f_\text{c} -\nicefrac{\Delta f}{2}, f_\text{c} +\nicefrac{\Delta f}{2}].
   \end{cases}
\end{equation}
Here $P_\text{n}$ is the total noise power measured at the end of the microwire with a calibrated diode and the resistance $R$ is set to \SI{50}{\ohm}. In the following, the noise frequency window is fixed to $\Delta f=\SI{50}{\MHz}$ with a center frequency $f_\text{c} = f_\text{NV} = \SI{2.87}{\GHz}$, which corresponds to the zero-field splitting between the electron spin sublevels $m_s=0$ and $m_s=\pm 1$ in the ground state of the NV defect [Fig.~\ref{fig:setup}(a)]. All the experiments are performed on a single NV defect localized at a distance $d = \SI[multi-part-units=single]{28\pm4}{\micro\meter}$ from the edge of the microwire without applying external static magnetic fields.\\

\begin{figure}[t]
  \centering
  \includegraphics[scale=1]{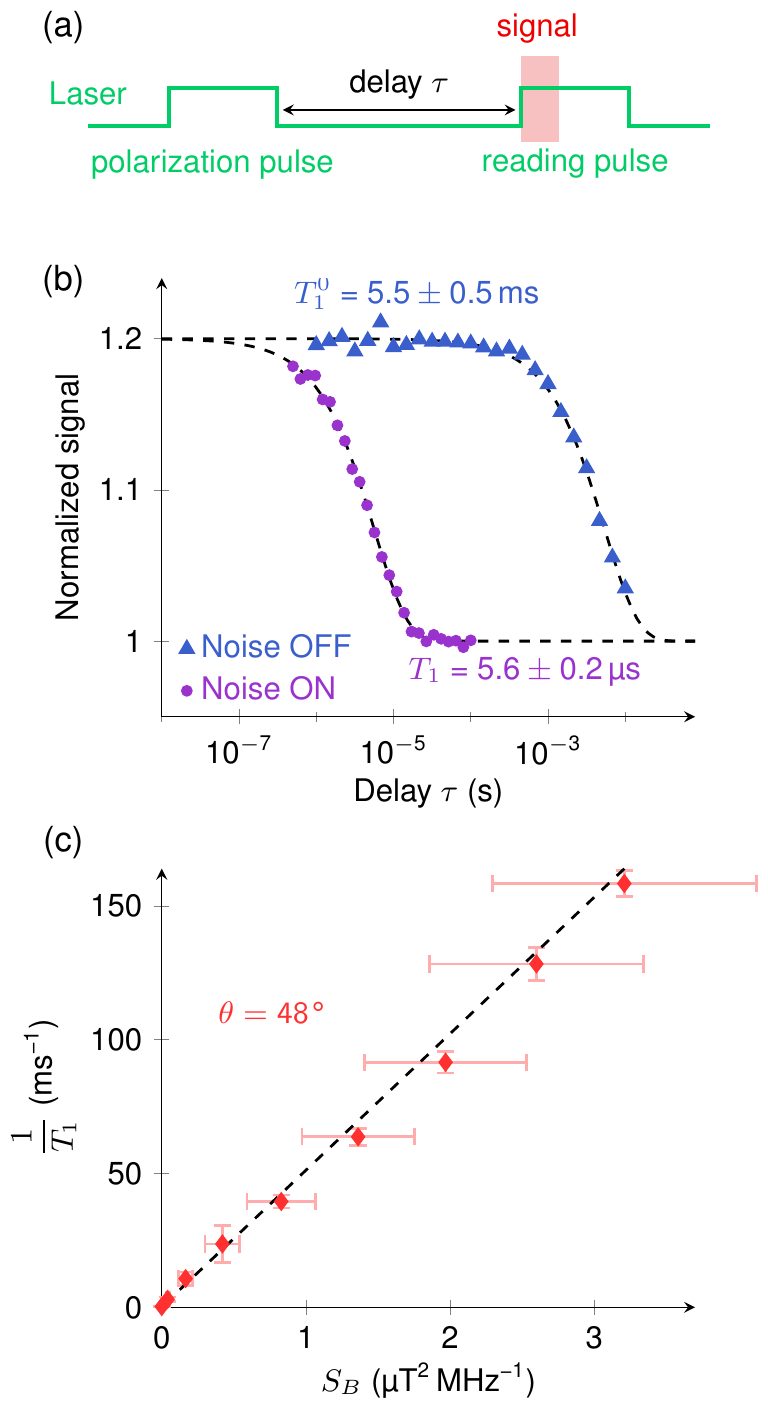}
  \caption{(a) Experimental sequence used to measure the longitudinal spin relaxation time $T_1$. The duration of the laser pulses is \SI{3}{\micro\second}. The spin-dependent PL signal is integrated in a detection window corresponding to the first \SI{300}{\nano\second} of the readout laser pulse and normalized with the value measured for $\tau \gg T_1$. (b) NV spin relaxation curves recorded in the absence (blue triangle) and presence (purple circles) of a magnetic noise with a frequency window $\Delta f=\SI{50}{\MHz}$, a center frequency $f_\text{c} = f_\text{NV} = \SI{2.87}{\GHz}$ and a field spectral density at the NV center position around \SI{3}{\micro\tesla\squared\per\mega\Hz}. The distance between the NV defect and the copper microwire is $d = \SI[multi-part-units=single]{28\pm4}{\micro\meter}$. (c) Dependence of the spin relaxation rate with the field spectral density. The black dashed line is a linear fit with Eq.~(\ref{eq:invT1_vs_SBperp}), leading to $\theta=\ang{48}$.}
  \label{fig:T1_vs_noise}
\end{figure}

We first calibrate precisely the impact of magnetic noise on the longitudinal spin relaxation time $T_1$ of the NV defect by using the measurement sequence shown in Fig.~\ref{fig:T1_vs_noise}(a). After initialization in the $m_s=0$ spin sublevel by optical pumping with a first green laser pulse, the NV defect relaxes in the dark during a variable time~$\tau$ through the two-way transition rate $\Gamma_1$ that couples the $m_s=0$ and $m_s=\pm 1$ sublevels [Fig.~\ref{fig:setup}(a)]. A second laser pulse is then used to probe the final spin population in $m_s=0$ by recording the spin-dependent PL signal, which decays exponentially with a characteristic lifetime $T_1=\nicefrac{1}{3\Gamma_1}$~\cite{tetienneSpinRelaxometrySingle2013}. Without applying magnetic noise, we measure an intrinsic spin relaxation time $T_1^0 = \SI[multi-part-units=single]{5.5 \pm 0.5}{\milli\second}$, a value commonly obtained for single NV defects hosted in ultrapure bulk diamond crystals at room temperature~\cite{jarmolaTemperatureMagneticFieldDependentLongitudinal2012}. Turning on the magnetic noise at the maximum available power, which corresponds to a field spectral density $S_B^\text{max} \sim \SI{3}{\micro\tesla\squared\per\mega\Hz}$ at the position of the studied NV defect, the spin relaxation time drops by three orders of magnitude, reaching $T_1 = \SI[multi-part-units=single]{5.6 \pm 0.2}{\micro\second}$ [Fig.~\ref{fig:T1_vs_noise}(b)]. 

%integrating the spin-dependent PL signal within a detection window corresponding to the first $300$~ns of the readout laser pulse. , which decays exponentially with $\tau$ decay with a characteristic relaxation time $T_1=1/3\Gamma_1$ where 
In the presence of a magnetic noise with a spectral component at the NV defect's electron spin resonance frequency $f_\text{NV}$, the longitudinal spin relaxation rate is given by~\cite{slichterPrinciplesMagneticResonance1990}
\begin{equation}
  \label{eq:invT1_vs_SBperp}
  \frac{1}{T_1} = \frac{1}{T_1^0} + 3 \gamma^2 S_{B_\perp}(f_\text{NV}),
\end{equation}
where $\gamma = \SI{28}{\GHz \per \tesla}$ is the electron gyromagnetic ratio and $S_{B_\perp}(f_\text{NV})$ is the field spectral density perpendicular to the NV defect quantization axis. In our experiment, the direction of the fluctuating Oersted field is fixed with respect to the microwire such that
\begin{equation}
  \label{eq:def_S_Bperp}
  S_{B_\perp}(f_\text{NV}) = S_B(f_\text{NV})\sin^2\theta \ ,
\end{equation}
where $\theta$ is the angle between the field direction and the quantization axis of the NV defect, as illustrated in Fig.~\ref{fig:setup}(a). The plot in Fig.~\ref{fig:T1_vs_noise}(c) shows the evolution of the spin relaxation rate with the field spectral density~$S_B$, which is tuned by varying the total noise power~$P_\text{n}$ sent through the microwire [see~Eq.~(\ref{eq:S_B})]. Despite large error bars originating from the uncertainty on the distance $d$ between the NV center and the microwire, the data displays the expected linear behavior. Data fitting with Eq.~(\ref{eq:invT1_vs_SBperp}) using the angle $\theta$ as only free parameter, leads to $\theta = \ang{48}$, a value compatible with the geometry of our setup. Next, this calibration measurement is used to tune $T_1$ in a quantitative fashion by varying $P_\text{n}$.\\

\begin{figure}[t]
  \centering
  \includegraphics[scale=1]{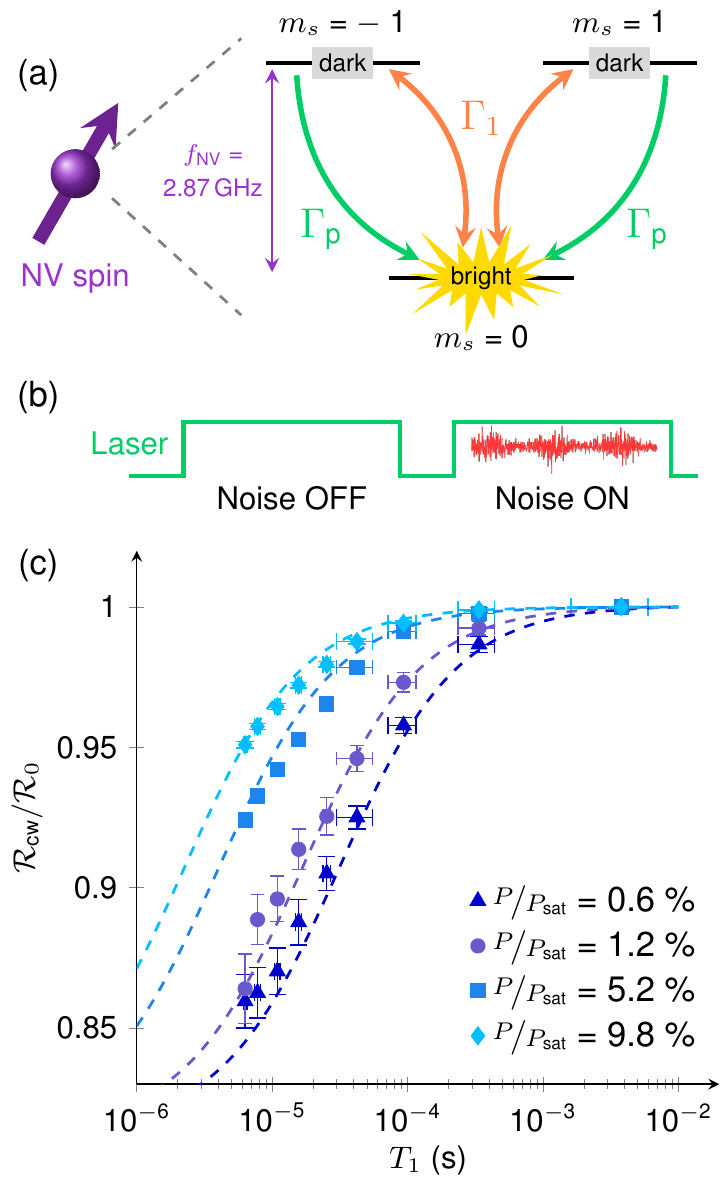}
  \caption{(a) Sketch of the three-level model used to describe the NV defect. (b) Measurement procedure used to infer the ratio between the PL signals in the presence ($\mathcal{R}_{\rm cw}$) and absence ($\mathcal{R}_{0}$) of magnetic noise. The two laser pulses have a duration of $1$~ms and are separated by \SI{50}{\micro\second}. (c) Dependence of the ratio $\mathcal{R}_{\rm cw}/\mathcal{R}_{0}$ with the spin relaxation time $T_1$ for different optical excitation powers. The dashed lines show the predictions of the closed three-level model [Eq.~(\ref{eq:PL_rate_full})] using $\beta=0.72$.}
  \label{fig:2_level_model}
\end{figure}

We now investigate the effect of magnetic noise on the PL signal emitted under continuous optical illumination. To this end, the NV defect is described by a closed three-level system corresponding to the $ m_s=0$ and $m_s = \pm 1$ spin sublevels in the ground state, as sketched in Fig.~\ref{fig:2_level_model}(a). Under optical excitation with a power $\mathcal{P}$, the steady state spin populations $(n_0^\text{st},n_{+1}^\text{st},n_{-1}^\text{st}$) result from the competition between the two-way transition rate $\Gamma_1$ and optically-induced spin polarization in the $m_s=0$ spin sublevel with a rate $\Gamma_\text{p}$ that depends on the rate of optical cycles. Denoting $\mathcal{P}_\text{sat}$ the saturation power of the optical transition, the spin polarization rate is expressed as~\cite{dreauAvoidingPowerBroadening2011}
\begin{equation}
\displaystyle \Gamma_\text{p} = \Gamma_\text{p}^\infty\frac{\mathcal{P}}{\mathcal{P}+\mathcal{P}_\text{sat}} \ ,
\end{equation}
where $\Gamma_\text{p}^\infty \simeq \SI{5e6}{\per\second}$ is the polarization rate at saturation, which is fixed by the lifetime of the metastable state ($\sim \SI{200}{\nano\second}$)~\cite{Robledo_2011} involved in the optically-induced spin polarization process.

Using this three-level model, the PL rate $\mathcal{R}_\text{cw}$ of the NV defect can be written as
\begin{equation}
  \label{eq:1}
  \mathcal{R}_\text{cw} = \mathcal{R}_0  n_0^\text{st} + \mathcal{R}_{\pm 1} (n_{+1}^\text{st} + n_{-1}^\text{st}) \ ,
\end{equation}
where $\mathcal{R}_0$ and  $\mathcal{R}_{\pm 1}$ are the PL rates associated with the spin populations in $m_s=0$ and $m_s=\pm1$, respectively. The spin-dependent PL response of the NV defect is then phenomenologically introduced by considering that $\mathcal{R}_1=$~$\beta \mathcal{R}_0$, with $\beta<1$. By solving the rate equations at the steady state in the three-level system, the PL rate finally reads
\begin{equation}
  \label{eq:PL_rate_full}
  \mathcal{R}_\text{cw} = \mathcal{R}_0  \frac{(1+2\beta)\Gamma_1 + \Gamma_\text{p}}{3\Gamma_1 + \Gamma_\text{p}},
\end{equation}

with
\begin{equation}
\mathcal{R}_0=\mathcal{R}_0^\infty\frac{\mathcal{P}}{\mathcal{P}+\mathcal{P}_\text{sat}}.
\end{equation}
Here $\mathcal{R}_0^\infty$ denotes the PL rate at saturation of the optical transition. In our experiments $\mathcal{R}_0^\infty\sim \SI{1e5}{\per\second}$ and $\mathcal{P}_\text{sat}\sim \SI{300}{\micro\watt}$. In the limit $\Gamma_1\ll \Gamma_\text{p}$, the NV defect is efficiently polarized in the $m_s=0$ spin sublevel, leading to a maximal PL rate~$\mathcal{R}_0$. Conversely, when $\Gamma_1\gg \Gamma_\text{p}$, spin polarization becomes inefficient and the PL signal drops to $\mathcal{R}_0\nicefrac{(1+2\beta)}{3}$.

The variations of the PL signal with magnetic noise is measured by using a sequence of two \si{\milli\second}-long laser pulses, as sketched in Fig~\ref{fig:2_level_model}(b). During the first pulse, no magnetic noise is applied such that $T_1=T_1^0$. In this case, the condition $\Gamma_1\ll \Gamma_\text{p}$ is always fulfilled in our experiment, leading to a reference PL rate $\mathcal{R}_0$. During the second pulse, the magnetic noise is switched on at different noise powers $P_\text{n}$, which leads to a reduction of the spin relaxation time $T_1=\nicefrac{1}{3\Gamma_1}$ according to the calibration measurement shown in Fig.~\ref{fig:T1_vs_noise}(c). The PL rate measured in the presence of noise $\mathcal{R}_\text{cw}$ is normalized to $\mathcal{R}_0$ and plotted as a function of $T_1$ for various optical excitation powers in Fig~\ref{fig:2_level_model}(b). As expected, a significant drop of the PL signal is observed when $T_1$ becomes shorter. Moreover, the characteristic value of $T_1$ at which the PL starts to drop is reduced when the optical power, and therefore the spin polarization rate $\Gamma_\text{p}$, is increased. The whole set of data is in good agreement with the predictions of the three-level model for $\beta=0.72$ [dashed lines in Fig.~\ref{fig:T1_vs_noise}(c)]. A reduction of $T_1$ therefore results in a quenching of the PL signal under continuous optical illumination, providing a simple mechanism to detect magnetic noise. Moreover, our results indicate that tiny deviations from $T_1^0$ can only be detected for low excitation power.

To estimate the sensitivity of the measurement, we consider a single NV defect in the absence of magnetic field noise, \textit{i.e.} $T_1=T_1^0$, and we apply an infinitesimal field spectral density $\delta S_{B_\perp}(f_\text{NV})$. This magnetic noise results in a reduction of the electron spin relaxation time by $\delta T_1=3(T_1^0 \gamma)^2 \delta S_{B_\perp}(f_\text{NV})$, which is converted into a variation $\delta \mathcal{N}$ of the number of detected photons. Assuming that the detection is limited by photon shot noise $\delta \mathcal{N}_\text{s}$, the signal to noise ratio (SNR) reads
\begin{equation}
  \label{eq:expr_snr}
  \mathrm{SNR} = \frac{\delta \mathcal{N}}{\delta \mathcal{N}_\text{s}}=\frac{ \left|\frac{\partial \mathcal{R}_\text{cw}}{\partial T_1}\right| \delta T_1 \Delta t}{\sqrt{\mathcal{R}_\text{cw} \Delta t}},
\end{equation}
where $\Delta t$ is the acquisition time. Inserting Eq.~(\ref{eq:PL_rate_full}) into Eq.~(\ref{eq:expr_snr}), the SNR is maximized for an optical power $\mathcal{P}_\text{opt}$ which verifies
\begin{equation}
  \label{eq:P_opt}
  \frac{\mathcal{P}_\text{opt}}{\mathcal{P}_\text{opt}+ \mathcal{P}_\text{sat}} = \frac{1+\sqrt{2(1+\beta)}}{\Gamma_\text{p}^\infty \ T_1^0}.
\end{equation}
Using this optimized optical power, the sensitivity $\eta_\text{cw}$ is then defined as the minimal variation $\delta S_{B_\perp}(f_\text{NV})$ than can be detected for a SNR equal to unity, leading to  
 \begin{equation}
  \label{eq:delta_t1}
  \displaystyle  \eta_\text{cw}=\delta S_{B_\perp}(f_\text{NV})\sqrt{\Delta t}=\frac{\Theta_\text{cw}}{\gamma^2 \sqrt{T_1^0}},
  \end{equation}
with
 \begin{equation}
 \Theta_\text{cw}=\frac{\sqrt{4+2\beta +\alpha}}{2\sqrt{3}(1-\beta)} \left(\frac{2+\alpha}{1+\alpha}\right)^{\frac{3}{2}} \sqrt{\frac{\Gamma_\text{p}^\infty}{\mathcal{R}_0^\infty}}, 
 \end{equation}
where we have set $\alpha= \sqrt{2(1+\beta)}$. Using $\beta=0.72$, we obtain $\Theta_\text{cw} \approx 40$, leading to a shot-noise limited sensitivity $\eta_\text{cw}\sim \SI{1}{\micro\tesla\squared{\mega\Hz}^{-1}\per\sqrt{\rm Hz}}$.

In order to better qualify the performances of this method, we compare it to a more commonly used relaxometry technique~\cite{Steinert2013,schmid-lorchRelaxometryDephasingImaging2015,TetienneNanolett2016}, which consists in repeating the pulse sequence shown in Fig.~\ref{fig:T1_vs_noise}(a) with a fixed time delay~$\tau$. For such a single-$\tau$ sensing procedure, the spin readout contrast is optimized for laser pulses with a typical duration $T_\text{L}\sim \SI{300}{\nano\second}$~\cite{dreauAvoidingPowerBroadening2011} and an optical power saturating the transition, \textit{i.e.} $\mathcal{P}=\mathcal{P}_\text{sat}$. In this case, the rate of detected photons is limited by the duty cycle of the laser pulses. Within the simplified three-level model used in this work, the resulting PL rate can be expressed as~\cite{tetienneSpinRelaxometrySingle2013}
 \begin{equation}
   \mathcal{R}_\text{pulse}=\frac{\mathcal{R}_0^{\infty}T_\text{L}}{2\tau}
   \left[1+\frac{2(1-\beta)}{1+2\beta}e^{-\tau/T_1}\right],
  \end{equation}
where we assume that $T_\text{L}\ll \tau$. Using a single-$\tau$ sensing procedure, the SNR is optimized for a delay $\tau_\text{opt}=\nicefrac{T_1^0}{2}$, leading to a shot noise limited sensitivity 
 \begin{equation}
  \displaystyle  \eta_\text{pulse}=\delta S_{B_\perp}(f_\text{NV})\sqrt{\Delta t}=\frac{\Theta_\text{pulse}}{\gamma^2 \sqrt{T_1^0}},
  \end{equation}
where 
 \begin{equation}
 \displaystyle \Theta_\text{pulse}=\frac{1+2\beta}{1-\beta}\sqrt{\frac{e}{3\mathcal{R}_0^{\infty}T_\text{L}}} \approx 50.
  \end{equation}

  \vspace*{0em}
  
This analysis indicates that the two methods are equivalent in terms of sensitivity, $\eta_\text{cw}\sim \eta_\text{pulse}$. However, besides being practically very simple to implement, relaxometry based on PL quenching under continuous optical illumination offers optimal performances at low optical excitation power, which could be advantageous for applications in cellular environments~\cite{Steinert2013,Rendler2017} and might enable to mitigate optically-induced charge-state conversion of the NV defect~\cite{Aslam_2013}. In addition, this relaxometry method is well suited for studying compensated magnetic materials, such as antiferromagnets, in which static stray fields are vanishingly small but thermal magnons producing magnetic noise are amply present~\cite{flebusProposalDynamicImaging2018}. This capability is best exemplified by the recent all-optical imaging of non collinear spin textures in synthetic antiferromagnets by use of noise-induced PL quenching~\cite{fincoImagingNoncollinearAntiferromagnetic2020}. As such, it is a complementary imaging mechanism to the one relying on PL quenching induced by large off-axis magnetic fields~\cite{tetienne_magnetic-field-dependent_2012}, which can be used for studying the physics of spin textures in ferromagnetic systems~\cite{gross_skyrmion_2018, akhtar_current-induced_2019}.

\begin{acknowledgements}
We thank Farah Amar (C2N) and Paul Crozat (C2N) for helpful hints on the design of the experiment. This research has received funding from the European Research Council (ERC) under Grant Agreement No. 866267 (EXAFONIS) and the French Agence Nationale de la Recherche through the project TOPSKY (Grant No. ANR-17-CE24-0025). A.F. acknowledges financial support from the EU Horizon 2020 Research and Innovation program under the Marie Sklodowska-Curie Grant Agreement No. 846597 (DIMAF).
\end{acknowledgements}

\bibliography{NV_noise_paper.bib}
\bibliographystyle{apsrev4-1}

\end{document}